\documentclass[11pt,a4paper]{amsart}
\usepackage{amsmath}
\usepackage{amssymb}
\usepackage{amsfonts}
\usepackage[toc,page]{appendix}
\usepackage{hyperref}

\def\={\ =\ }

\newcommand{\be}{\begin{equation}}
\newcommand{\ee}{\end{equation}}
\newcommand{\beq}{\begin{equation}}
\newcommand{\eeq}{\end{equation}}
\newcommand{\bea}{\begin{eqnarray}}
\newcommand{\eea}{\end{eqnarray}}

\def\ba{\begin{eqnarray}}
\def\ea{\end{eqnarray}}

\theoremstyle{plain}

\numberwithin{equation}{section}

\input{tcilatex}

\begin{document}
\title[\hfill\ \ ]{Chern-Simons theory encoded on a spin chain}
\author{David P\'{e}rez-Garc\'{\i}a}
\address{Departamento de An\'{a}lisis Matem\'{a}tico and Instituto de Matem%
\'{a}tica Interdisciplinar, Facultad de Ciencias Matem\'{a}ticas,
Universidad Complutense de Madrid, 28040 Madrid, Spain}
\email{dperez@mat.ucm.es}
\author{Miguel Tierz}
\address{Departamento de Matem\'{a}tica, Grupo de F\'{\i}sica Matem\'{a}%
tica, Faculdade de Ci\^{e}ncias, Universidade de Lisboa, Campo Grande, Edif%
\'{\i}cio C6, 1749-016 Lisboa, Portugal.}
\email{tierz@fc.ul.pt}
\address{Departamento de An\'{a}lisis Matem\'{a}tico, Facultad de Ciencias
Matem\'{a}ticas, Universidad Complutense de Madrid, 28040 Madrid, Spain}
\email{tierz@mat.ucm.es}

\begin{abstract}
We construct a 1d spin chain Hamiltonian with generic interactions and prove
that the thermal correlation functions of the model admit an explicit random
matrix representation. As an application of the result, we show how the
observables of $U(N)$ Chern-Simons theory on $S^{3}$ can be reproduced with
the thermal correlation functions of the 1d spin chain, which is of the XX
type, with a suitable choice of exponentially decaying interactions between
infinitely many neighbours. We show that for this model, the correlation
functions of the spin chain at a finite temperature $\beta =1$ give the
Chern-Simons partition function, quantum dimensions and the full topological 
$S$-matrix.
\end{abstract}

\maketitle

\section{Introduction}

The study of topological field theories has been a subject of continued
interest in the last decades (see \cite{topo-rev} for an early review).
Chief among topological field theories is Chern-Simons theory, due to its
relevance in the study of three dimensional topology and in knot theory \cite%
{Witten:1988hf}, its role as the effective field theory of the fractional
quantum Hall effect \cite{QHeff} and, more recently, its appearance also in
the study of topological quantum computation \cite{topq,topbook,BP} and in
the description of topological strings \cite{Marino:2004uf}.

In recent years, motivated by the realm of quantum computation, a number of
proposals have been devised to simulate different quantum field theories
(QFT), including non Abelian gauge theories \cite{QFTions}-\cite{ZCR2}. In a
somehow related but different direction, in \cite{DM}, we introduced a
connection, illustrated for the case of low-energy QCD, between some QFT and
1D spin systems, based on the existence of a random matrix description. This
allowed us to relate crucial properties of the QFT with physically
meaningful properties of the 1D system.

In this paper, following that path, we introduce a new spin chain
Hamiltonian, expected to be in the same universality class as the
XX-Hamiltonian, whose thermal correlation functions correspond exactly with
the main observables of Chern-Simons theory, including knot polynomials. We
will also discuss how a finite chain approximates these observables with an
exponentially small error in the size of the chain and how finite chains
models may describe Chern-Simons theory coupled with matter in the
fundamental representation.

Let us first remind the basics of Chern-Simons theory, which is a three
dimensional gauge theory with a simply connected and compact non-Abelian Lie
group $G$ and the Chern-Simons action, given by \cite{Witten:1988hf}%
\begin{equation}
S_{\mathrm{CS}}(A)={\frac{k}{4\pi }}\int_{M}\mathrm{Tr}(A\wedge dA+{\frac{2}{%
3}}A\wedge A\wedge A),  \label{cs}
\end{equation}%
where $\mathrm{Tr}$ is the trace in the fundamental representation, $A$ is
the connection, a 1-form valued on the corresponding Lie algebra, and $k\in 
\mathbb{Z}$ is the level. The manifold $M$ is a three dimensional compact
manifold which, in this work, will be chosen to be $S^{3}$. The $q$%
-parameter of Chern-Simons theory is defined in terms of the level $k$ by $%
q=\exp \left( 2\pi i/(k+h)\right) $ where $h$ is the dual Coxeter number of
the Lie algebra, which is just $h=N$ in this paper since we choose $G=U(N)$.
In the random matrix theory description of Chern-Simons theory that we will
employ, $q$ is treated as a real number $q=\exp (-\gamma )$ (with $\gamma >0$%
) and the Chern-Simons results are recovered at the end by making the
identification $\gamma =2\pi i/(k+h)$. The original study of knot polynomial
invariants by Jones also considers $q$ real \cite{Jones} although it is
possible to extend it to the case of roots of unity, in which case the
necessary representation theory of the Hecke algebra is more delicate to
deal with.

Let us introduce now a succinct reminder of the solution of Chern-Simons
theory. In \cite{Witten:1988hf}, a concrete description of the
(finite-dimensional) Hilbert space of the theory was given: it is the space
of conformal blocks of a Wess-Zumino-Witten (WZW) model \cite{CFT} on $%
\Sigma $ with gauge group $G$ and level $k$. Here $\Sigma $ denotes the
Riemann surface which is the common boundary in the Heegard splitting of the
manifold $M$ \cite{knot}. In the case studied here, $M=S^{3}$, the Riemann
surface is a torus and for $\Sigma =T^{2}$, the space of conformal blocks is
in one to one correspondence with the integrable representations of the
affine Lie algebra associated to $G$ at level $k$ \cite%
{Witten:1988hf,Marino:2004uf}. We recall that a representation given by a
highest weight $\Lambda $ is integrable if $\rho +\Lambda ,$ where $\rho $
denotes the Weyl vector, is in the fundamental chamber $\mathcal{F}_{l}$ ,
with $l=k+N$ in our case. Definitions of the Weyl vector, fundamental
chamber and all the other necessary concepts in representation theory are
given in the Appendix, see \cite{CFT} for a textbook treatment.

The states in the Hilbert state of the torus $\mathcal{H}(T^{2})$ are
denoted by $\left\vert \rho +\Lambda \right\rangle $ and, as we have just
explained, $\rho +\Lambda \in \mathcal{F}_{l}$ , since it is an integrable
representation of the WZW model at level $k$. We denote these states by $%
\left\vert \lambda \right\rangle $ where $\lambda $ is the representation
associated to $\Lambda $. This association is also made explicit in the
Appendix. The state $\left\vert \rho \right\rangle $ is denoted by $%
\left\vert 0\right\rangle $ and the states can be chosen to be orthonormal 
\cite{Witten:1988hf,Marino:2004uf} and therefore $\left\langle \lambda
^{\prime }\right\vert \left. \lambda \right\rangle =\delta _{\lambda
,\lambda ^{\prime }}$.

In addition, there is a special class of homeomorphisms of $T^{2}$ that have
a simple expression as operators in $\mathcal{H}(T^{2})$: the $S$ and $T$
transformations, which are the generators of the $SL(2,%
\mathbb{Z}
)$ group. The matrix elements of these transformations, $T_{pp^{\prime }}$
and $S_{pp^{\prime }}$, have a very explicit expression in the basis of the
integrable representations, but these will not be needed here. What is
relevant here is that this non-perturbative exact solution of Chern-Simons
theory computes the observables of the theory, giving \cite%
{Witten:1988hf,Marino:2004uf}%
\begin{equation*}
Z\left( S^{3}\right) =\langle \,0\rvert \,{}S{}\,\lvert \,0\,\rangle =S_{00}.
\end{equation*}%
This is generalized to obtain the result of the path integral in $S^{3}$
with the insertion of knots and links. Considering a solid torus where a
Wilson line in representation $\lambda $ has been inserted (by inserting a
Wilson loop $\mathcal{O}_{\lambda }=\mathrm{Tr}_{\lambda }U$ in the
representation $\lambda $ in the path integral), then the corresponding
state is$\,\lvert \lambda \rangle $. Gluing this to an empty solid torus
after an S transformation, one obtains the unknot on $S^{3}$%
\begin{equation}
Z\left( S^{3},\mathcal{O}_{\lambda }\right) =\langle \,0\rvert
\,{}S{}\,\lvert \,\lambda \,\rangle =S_{0\lambda }.  \label{Un}
\end{equation}%
The normalization of (\ref{Un}) is the quantum dimension of the
representation%
\begin{equation*}
\dim _{q}\lambda =\frac{\langle \,0\rvert \,{}S{}\,\lvert \,\lambda
\,\rangle }{\langle \,0\rvert \,{}S{}\,\lvert \,0\,\rangle }=\frac{%
S_{0\lambda }}{S_{00}}.
\end{equation*}%
The link formed by two unknots in representations $\lambda $ and $\mu $ is
the Hopf link, given by%
\begin{equation*}
W_{\mu \lambda }=\frac{\langle \,\mu \rvert \,{}S{}\,\lvert \,\lambda
\,\rangle }{\langle \,0\rvert \,{}S{}\,\lvert \,0\,\rangle }=\frac{S_{\mu
\lambda }}{S_{00}}.
\end{equation*}%
These observables are (quantum) topological invariants of the manifold \cite%
{Turaev} and have an important interpretation in terms of the properties
(fusion and braiding) of the quasiparticles of a topological quantum field
theory. In particular, the elements of the S-matrix are closely related to
quasiparticle braiding since $S_{ab}$ is equal to the amplitude for creating 
$a\overline{a}$ and $b\overline{b}$ pairs, braiding $a$ and $b$, and
annihilating again in pair. In addition, as shown by Verlinde's formula, the 
$S$-matrix not only contains information about braiding, but also about
fusion \cite{topq}. The quantum dimension of an anyonic species is a measure
for the effective number of degrees of freedom of the internal Hilbert space
of the corresponding particle type \cite{Bais}.

\subsection{The two main results}

The first main result of this work is an integral representation of the
random matrix type, for the thermal correlation function, defined as: 
\begin{equation}
F_{j_{1},\dots ,j_{K};l_{1},\dots ,l_{K}}(\beta )=\langle \,\Uparrow
\,\rvert \,{}\sigma _{j_{1}}^{+}\cdots \sigma _{j_{K}}^{+}\mathrm{e}^{-\beta 
\hat{H}}\sigma _{l_{1}}^{-}\cdots \sigma _{l_{K}}^{-}{}\,\lvert \,\Uparrow
\,\rangle .  \label{F-def}
\end{equation}%
where$\lvert \,\Uparrow \,\rangle $ denotes a ferromagnetic state,
characterized by having all of its spins up $\lvert \Uparrow \,\rangle
=\otimes _{i}\lvert \uparrow \rangle $ which satisfies $\sigma
_{k}^{+}{}\,\lvert \,\Uparrow \,\rangle =0$ for all $k,$ and the state is
also normalized $\langle \,\Uparrow \,\mid \,\Uparrow \,\rangle =1$. The
operators inserted in (\ref{F-def}) are string operators, which in terms of
the Pauli spin operators are $\sigma _{i}^{\pm }=\left( \sigma _{i}^{x}\pm
i\sigma _{i}^{y}\right) /2.$ We consider a generic long-range 1d spin
Hamiltonian%
\begin{equation}
\hat{H}_{\mathrm{Gen}}=-\sum_{i}\dsum\limits_{j\in \mathbb{Z}}a_{j}\left(
\sigma _{i}^{-}\otimes \sigma _{i+j}^{+}\right) +\frac{h}{2}\sum_{i}(\sigma
_{i}^{z}-\mathbb{I}),  \label{Hgen-inf}
\end{equation}%
where the $a_{j}$ denote arbitrary real coefficients which at least decay as 
$a_{j}\sim j^{-1-\eta }$ with $\eta >0$. For the application to Chern-Simons
theory only exponential decay will be necessary. Our first new result will
show that the thermal correlator (\ref{F-def}) is given by a unitary random
matrix model 
\begin{eqnarray}
F_{j_{1},\dots ,j_{K};l_{1},\dots ,l_{K}}(\beta ) &=&\frac{\mathrm{e}^{\beta
hN}}{(2\pi )^{N}N!}\int\limits_{-\pi }^{\pi }d\varphi _{1}\!\cdots
\!\!\int\limits_{-\pi }^{\pi }d\varphi _{N}\prod_{1\leq j<k\leq N}\left\vert
e^{i\varphi _{k}}-e^{i\varphi _{j}}\right\vert ^{2}\left(
\prod_{j=1}^{N}f_{\beta }(e^{i\varphi _{j}})\right)   \notag \\
&&\times \overline{\hat{s}_{\mu }\left( e^{i\varphi _{1}},\dots ,e^{i\varphi
_{N}}\right) }\hat{s}_{\lambda }\left( e^{i\varphi _{1}},\dots ,e^{i\varphi
_{N}}\right) ,  \label{RMTrep}
\end{eqnarray}%
where the weight function $f_{\beta }(e^{i\varphi })$ of the random matrix
ensemble is 
\begin{equation}
f_{\beta }\left( e^{i\varphi }\right) =f_{0}\left( e^{i\varphi }\right) \exp
\left( \beta \sum_{k\in 
\mathbb{Z}
}a_{k}e^{ik\varphi }\right) .
\end{equation}%
The second main result of this work follows form a very specific
particularization of the first main result and consists in showing that all
the elements $S_{\lambda \mu }$ can be recovered as thermal (with $\beta =1$%
) correlation functions of a new 1D Hamiltonian with exponentially decaying
interactions, namely:%
\begin{equation}
S_{\lambda \mu }=\langle \,\Uparrow \,\rvert \,{}\sigma _{j_{1}}^{+}\cdots
\sigma _{j_{N}}^{+}\mathrm{e}^{-\hat{H}_{CS}}{}\,\sigma _{l_{1}}^{-}\cdots
\sigma _{l_{N}}^{-}\lvert \,\Uparrow \,\rangle   \label{main-eq}
\end{equation}

where%
\begin{equation}
\hat{H}_{CS}=-\sum_{i,k\in 
\mathbb{Z}
_{+}}\frac{(-1)^{k+1}}{2k\sinh \left( \frac{k\gamma }{2}\right) }\left(
\sigma _{i}^{-}\otimes \sigma _{i+k}^{+}+\sigma _{i}^{-}\otimes \sigma
_{i-k}^{+}\right) +\frac{h}{2}\sum_{i}(\sigma _{i}^{z}-\mathbb{I}),
\label{CS-H}
\end{equation}%
$S_{\lambda ,\mu }$ are the entries of the $S$-matrix of Chern-Simons theory
on $S^{3}$ and gauge group $U(N)$, again $\lvert \,\Uparrow \,\rangle $
denotes a ferromagnetic state, characterized by having all of its spins up
and $q=\mathrm{e}^{-\gamma }$, since the matrix model formulation that we
employ allows to work with the analytical continuation of Chern-Simons
theory with $\gamma $ real, which is of relevance in topological string
theory \cite{Marino:2004uf} and $q$-deformed 2d Yang-Mills theory \cite{q}.
The Chern-Simons observables are recovered from the computation of the
matrix models, that we identify here with (\ref{main-eq}), by identifying $%
q=\exp \left( 2\pi i/(k+N)\right) $. The representations $\lambda $ and $\mu 
$ are described by Young tableaux \cite{Marino:2004uf} and are related to
the pattern of flipped spins by $\lambda _{r}=j_{r}-N+r$ and $\mu
_{r}=l_{r}-N+r$ \cite{DM}. There is freedom in the choice of the magnetic
field $h$, which only affects the overall normalization factor in the
thermal correlation functions. We shall discuss its role in the context of
the framing of the Chern-Simons theory. Note that the rank of the gauge
group coincides with the number of flipped spins in the thermal correlation
function of the spin chain. As particular cases we get the partition
function and the quantum dimensions in Chern-Simons theory:%
\begin{equation}
S_{00}=\,\langle ...,\uparrow ,\underset{N}{\underbrace{\downarrow
,...,\downarrow }}\rvert {}\mathrm{e}^{-\hat{H}_{CS}}\,\lvert \underset{N}{%
\underbrace{\downarrow ,...,\downarrow }},\uparrow ,...\rangle ,  \notag
\end{equation}%
\begin{equation*}
S_{0\lambda }=\langle \,\Uparrow \,\rvert \,{}\sigma _{j_{1}}^{+}\cdots
\sigma _{j_{K}}^{+}\mathrm{e}^{-\hat{H}_{CS}}\lvert \underset{N}{\underbrace{%
\downarrow ,...,\downarrow }},\uparrow ,...\rangle .
\end{equation*}

\section{Derivation}

We first prove that the thermal correlators (\ref{F-def}) associated to the
general Hamiltonian (\ref{Hgen-inf}) have the random matrix representation (%
\ref{RMTrep}). The proof will be rigorous by working with a finite chain of $%
N+1$ sites (and periodic boundary conditions). This is so because the
determinantal expression that will be given for the difference-differential
equation satisfied by (\ref{F-def}) will be then unique. The
multi-dimensional Riemann sum expression for such a determinant converges
then to the unitary matrix model (\ref{RMTrep})\ in the $N\rightarrow \infty 
$ limit, which, at the same time, proves that expression (\ref{F-def}) is
well defined in the thermodynamical limit.

The second main result is then a particularization of the first main result,
but a very specific one: the coefficients are chosen to be the Fourier
coefficients of the logarithm of a theta function. This makes the
corresponding generic matrix model to be the one that describes $U(K)$
Chern-Simons theory on $S^{3}$.

\subsection{Derivation of Main Result 1}

We begin by considering the XX Hamiltonian extended to admit generic
interactions, denoted by $a_{j}$, to all of its neighbours and on a
one-dimensional lattice consisting of $N+1$ sites labeled by elements of the
set $M\equiv \{0\leq k\leq N,k\in 
\mathbb{Z}
\}$, $N+1=0$ $(\mathrm{mod}$ $2)$. For simplicity of the presentation we
take $N+1$ to be odd and choose the middle site to be indexed by $i=0$. The
Hamiltonian is then%
\begin{equation}
\hat{H}_{\mathrm{Gen}}=-\sum_{i=0}^{N}\dsum\limits_{j=1}^{\left( N-1\right)
/2}a_{j}\left( \sigma _{i}^{-}\otimes \sigma _{i+j}^{+}+\sigma
_{i}^{-}\otimes \sigma _{i-j}^{+}\right) +\frac{h}{2}\sum_{i=0}^{N}(\sigma
_{i}^{z}-\mathbb{I}).  \label{Hgen}
\end{equation}%
Note that this is the finite chain, with $N+1$ sites, of the Hamiltonian (%
\ref{Hgen-inf}). We have the following commutation relations%
\begin{equation}
\lbrack \sigma _{j}^{+},\hat{H}]=-\sigma _{j}^{z}\sum_{i=0}^{\left(
N-1\right) /2}a_{i}(\sigma _{j+i}^{+}+\sigma _{j-i}^{+})-h\sigma _{j}^{+},
\end{equation}%
which, reasoning as in \cite{DM}, give for the $K=1$ correlation function%
\begin{equation*}
F_{jl}(\beta )\equiv \langle \,\Uparrow \,\rvert \,{}\sigma _{j}^{+}\mathrm{e%
}^{-\beta \hat{H}_{\mathrm{Gen}}}\sigma _{l}^{-}{}\,\lvert \,\Uparrow \text{ 
}\rangle \,,
\end{equation*}%
the following differential-difference equation%
\begin{equation}
\frac{d}{d\beta }F_{jl}(\beta )=\sum_{k=1}^{(N-1)/2}a_{k}F_{j+k,l}(\beta
)+\sum_{k=1}^{(N-1)/2}a_{k}F_{j-k,l}(\beta ).  \label{rw}
\end{equation}%
The generating function associated to these correlators $f_{\beta }\left(
\lambda \right) =\sum_{j=0}^{N}F_{jl}(\beta )\lambda ^{j}$ reads%
\begin{equation}
f_{\beta }\left( \lambda \right) =f_{0}\left( \lambda \right) \exp \left(
\beta \left( \sum_{k=1}^{(N-1)/2}a_{k}\lambda
^{k}+\sum_{k=1}^{(N-1)/2}a_{k}\lambda ^{-k}\right) \right) .  \label{gen}
\end{equation}%
In the $K>1$ case, the same procedure leads for the correlator, in the
simplest setting of the XX chain%
\begin{equation*}
G_{j_{1},\dots ,j_{K};l_{1},\dots ,l_{K}}(\beta )=\langle \,\Uparrow
\,\rvert \,{}\sigma _{j_{1}}^{+}\cdots \sigma _{j_{K}}^{+}\mathrm{e}^{-\beta 
\hat{H}_{XX}}{}\,\sigma _{l_{1}}^{-}\cdots \sigma _{l_{K}}^{-}\lvert
\,\Uparrow \,\rangle ,
\end{equation*}%
it simply satisfies a multi-index version of the difference-differential
equation for the $K=1$ case. In particular%
\begin{equation*}
\frac{d}{d\beta }G_{j_{1},\dots ,j_{K};l_{1},\dots ,l_{K}}(\beta )=\frac{1}{2%
}\sum_{l=1}^{K}\left( G_{j_{1},\dots ,j_{l+1},...,j_{K};l_{1},\dots
,l_{K}}(\beta )+G_{j_{1},\dots ,j_{l-1},...,j_{K};l_{1},\dots ,l_{K}}(\beta
)\right) 
\end{equation*}%
As pointed out in \cite{Bogo}, the solution to this equation, as can be
easily checked explicitly, can be written as a determinant whose entries
satisfy the $K=1$ differential-difference equation. The key point is that
the same result holds for the generalization considered here. In particular,
in our case, we have that, for $K\leq N$%
\begin{equation}
\frac{d}{d\beta }F_{j_{1},\dots ,j_{K};l_{1},\dots ,l_{K}}(\beta )=\frac{1}{2%
}\sum_{k=1}^{(N-1)/2}\sum_{l=1}^{K}a_{k}F_{j_{1},\dots
,j_{l}+k,...,j_{K};l_{1},\dots ,l_{K}}(\beta
)+\sum_{k=1}^{(N-1)/2}\sum_{l=1}^{K}a_{k}F_{j_{1},\dots
,j_{l}+k,...,j_{K};l_{1},\dots ,l_{K}}(\beta ),  \label{mult-index}
\end{equation}%
and the determinantal form for the correlator%
\begin{equation}
F_{j_{1},\dots ,j_{K};l_{1},\dots ,l_{K}}(\beta )=\det \left(
F_{j_{r}l_{s}}\right) _{r,s=1}^{K}  \label{detmin}
\end{equation}%
satisfies (\ref{mult-index}), if the $F_{j_{r}l_{s}}$ satisfy (\ref{rw}), as
can be easily checked explicitly. The Fourier coefficients of the generating
function contain the correlators. In the $K=1$ and for a finite $N$ chain,
they read\footnote{%
The $N+1$ is choosing the convention of Bogoliubov et al where of $N+1$
sites labeled by elements of the set $M\equiv \{0\leq k\leq N,k\in 
\mathbb{Z}
\}$, $N+1=0$ $(\mathrm{mod}$ $2)$ \cite{BC}. See also \cite{BC} for the
Riemann sum expression which corresponds to the XX Hamiltonian.}%
\begin{equation*}
F_{jl}(\beta )=\frac{1}{N+1}\sum_{s=0}^{N}\exp \left( \beta \left(
\sum_{k=1}^{(N-1)/2}a_{k}e^{ik\theta
_{s}}+\sum_{k=1}^{(N-1)/2}a_{k}e^{-ik\theta _{s}}\right) \right) e^{i\theta
_{s}(j-l)},
\end{equation*}%
where $\theta _{s}=\frac{2\pi }{N+1}(s-\frac{N}{2})$. Notice that this
extends the result in \cite{BC} to the case of arbitrary number of neighbour
interactions. Likewise, the multi-dimensional Fourier coefficient is then,
with the same expression for $\theta _{s}$ as above with $s=0,1,...N$, as in 
\cite{BC} 
\begin{eqnarray}
F_{j_{1},\dots ,j_{K};l_{1},\dots ,l_{K}}(\beta ) &=&\frac{1}{\left(
N+1\right) ^{K}}\sum_{N\geq s_{1}>s_{2}>...>s_{K}\geq 0}^{{}}\exp \left(
\beta \sum_{l=1}^{K}\left( \sum_{k=1}^{(N-1)/2}a_{k}e^{ik\theta
_{s_{l}}}+\sum_{k=1}^{(N-1)/2}a_{k}e^{-ik\theta _{s_{l}}}\right) \right)  
\notag \\
&&\times \prod_{1\leq j<k\leq K}\left\vert e^{i\theta _{k}}-e^{i\theta
_{j}}\right\vert ^{2}\overline{\hat{s}_{\mu }\left( e^{i\theta _{1}},\dots
,e^{i\theta _{K}}\right) }\hat{s}_{\lambda }\left( e^{i\theta _{1}},\dots
,e^{i\theta _{K}}\right) \text{.}  \label{F-discrete}
\end{eqnarray}%
This analytical expression can also be understood to follow immediately from
the Toeplitz minor identity for determinants of the type (\ref{detmin}),
given in \cite{BD}. In this discrete random matrix representation of the
correlator, $\hat{s}_{\lambda }\left( e^{i\varphi },\dots ,e^{i\varphi
}\right) $ is a Schur polynomial \cite{BD} and the representation $\lambda $
is indexed by a partition $\left\{ \lambda _{i}\right\} _{i=1}^{N}$ which is
related to the sequence $\left\{ j_{i}\right\} _{i=1}^{N}$\ in the spin
operators by $\lambda _{r}=j_{r}-N+r$ \cite{B}. The other sequence in the
correlator is correspondingly accounted for by the other Schur polynomial in
(\ref{F-discrete}).

By taking the slowest decay of the coefficients to be $a_{j}\sim j^{-1-\eta
} $ with $\eta >0$, the limit of an infinite spin chain $N\rightarrow \infty 
$ renders these Riemann sums into their corresponding integral form \cite%
{RS-rev}, leading to a $U(K)$ unitary random matrix ensemble%
\begin{equation}
\int\limits_{-\pi }^{\pi }d\varphi _{1}\!\cdots \!\!\int\limits_{-\pi }^{\pi
}d\varphi _{K}\prod_{1\leq j<k\leq K}\left\vert e^{i\varphi
_{k}}-e^{i\varphi _{j}}\right\vert ^{2}\left( \prod_{j=1}^{K}f_{\beta
}(e^{i\varphi _{j}})\right) \overline{\hat{s}_{\mu }\left( e^{i\varphi
_{1}},\dots ,e^{i\varphi _{K}}\right) }\hat{s}_{\lambda }\left( e^{i\varphi
_{1}},\dots ,e^{i\varphi _{K}}\right) ,  \label{matb}
\end{equation}

\subsection{Derivation of Main result 2.}

The next step is to recall from \cite{Okuda:2004mb,RT} that $U(N)$
Chern-Simons theory on $S^{3}$ can be described in terms of a unitary matrix
model. In particular, the matrix model is of the type (\ref{matb}) but with
a weight function 
\begin{equation}
f(\varphi )=\Theta _{3}({\,\mathrm{e}}\,^{{\,\mathrm{i}\,}\varphi
_{j}}|q)=\sum_{n=-\infty }^{\infty }\mathrm{q}^{n^{2}/2}\mathrm{e}%
^{in\varphi }\text{,}  \label{theta}
\end{equation}%
which is Jacobi's third theta function.

Hence, to reproduce Chern-Simons theory we have to consider the generalized
spin chain (\ref{Hgen}) in the specific setting where the resulting
generating function (\ref{gen}) turns out to be the theta function (\ref%
{theta}). The Fourier coefficients of $\ln \Theta _{3}({\,\mathrm{e}}\,^{{\,%
\mathrm{i}\,}\varphi _{j}}|q)$ can be easily obtained by using the product
form of (\ref{theta}) given by the Jacobi triple product identity%
\begin{equation}
\Theta _{3}({\,\mathrm{e}}\,^{{\,\mathrm{i}\,}\varphi
}|q)=\prod\limits_{j=1}^{\infty }\left( 1-q^{j}\right) \left( 1+q^{j-1/2}{%
\mathrm{e}}\,^{{\,\mathrm{i}\,}\varphi }\right) \left( 1+q^{j-1/2}{\mathrm{e}%
}\,^{{\,-\mathrm{i}\,}\varphi }\right) .  \label{triple}
\end{equation}%
Taking the logarithm leads to%
\begin{equation}
\log \Theta _{3}({\,\mathrm{e}}\,^{{\,\mathrm{i}\,}\varphi
}|q)=\sum_{j=1}^{\infty }\log \left( 1-q^{j}\right) +\sum_{j=1}^{\infty
}\left( \log \left( 1+q^{j-1/2}z\right) +\log \left( 1+q^{j-1/2}/z\right)
\right) .  \label{log}
\end{equation}%
The first term is the Fourier coefficient of order zero. We discuss its
relevance below and focus on the non-trivial part of (\ref{log}). Using the
Taylor expansion of $\log \left( 1+x\right) $, we have%
\begin{eqnarray*}
\sum_{j=1}^{\infty }\left( \log \left( 1+q^{j-1/2}z\right) +\log \left(
1+q^{j-1/2}/z\right) \right) &=&\sum_{j=1}^{\infty }\sum_{n\in 
\mathbb{Z}
,n\neq 0}^{\infty }\frac{\left( -1\right) ^{n+1}}{n}q^{\left( j-1/2\right)
n}z^{n} \\
&=&\sum_{n\in 
\mathbb{Z}
,n\neq 0}^{\infty }\frac{\left( -1\right) ^{n+1}}{n}z^{n}\sum_{j=1}^{\infty
}q^{n\left( j-1/2\right) } \\
&=&\sum_{n\in 
\mathbb{Z}
,n\neq 0}^{\infty }\frac{\left( -1\right) ^{n+1}q^{n/2}}{n(1-q^{n})}z^{n}.
\end{eqnarray*}%
Therefore, the Fourier coefficient $a_{n}$ of order $n$ ($n\in 
\mathbb{N}
$), reads%
\begin{equation}
a_{n}=a_{-n}=\frac{\left( -1\right) ^{n+1}q^{n/2}}{n(1-q^{n})},
\label{Fourier}
\end{equation}%
and we have that%
\begin{equation}
a_{k}=\frac{\left( -1\right) ^{k+1}}{2k\sinh \left( \frac{k\gamma }{2}%
\right) },  \label{decay}
\end{equation}%
with $k\in 
\mathbb{Z}
$ and with $q=\exp (-\gamma )$ the $q$-parameter of Chern-Simons theory. In
addition, we have to choose $\beta =1$ in (\ref{gen}) because plugging (\ref%
{decay}) in (\ref{gen}) gives $\Theta _{3}^{\beta }({\,\mathrm{e}}\,^{{\,%
\mathrm{i}\,}\varphi _{j}}|q)$. The orthogonality of the states fixes the
initial condition for $\beta =0$ to be $\delta _{j,l}$ and this implies that 
$g_{0}\left( \lambda \right) =1$. This finishes the proof of the main result.

Notice that the Fourier coefficient of order zero, which is $%
a_{0}=N\sum_{n=1}^{\infty }\ln (1-q^{n})$ has not been included. It only
contributes with a normalization factor and hence can be taken into account
by fixing the value of the external magnetic field. In any case, this choice
of normalization gives an exact correspondence with the unitary matrix model
with a theta function \cite{ST,RT}. Its partition function is explicitly
given below where we also explain that framing can be taken into account by
a proper choice of the magnetic field.

\section{Discussion}

A recent result \cite{BaikLiu} shows that the error obtained by
approximating (\ref{matb}) by its Riemann sum decreases exponentially with $%
L $. \ More concretely, the relative error is $\mathcal{O}\left(
e^{-c(L-N)}\right) $ as $L-N\rightarrow \infty $, even if $N$ also goes to $%
\infty $. This result holds for Toeplitz determinants of discrete measure
under very general conditions and it immediately applies to $S_{00}$. The
same result is expected to hold for $S_{\mu \lambda }$ because due to (\ref%
{main-eq}) and (\ref{matb}) we know it is the minor of a Toeplitz matrix 
\cite{BD}.\ These results suggest that the quantum topology of a manifold
can be described with a finite ring of spins interacting as explained above 
\footnote{%
The discussion on experimental accessibility made in \cite{DM} for the
XX-model applies also here.}.

The finite chain case is also of special interest in its own right because
the discretization of the matrix model also emerges naturally in the study
of Chern-Simons theory on $S^{2}\times S^{1}$ with matter in the fundamental
representation of the gauge group (vector matter). In recent work \cite%
{Jain:2013py}, it has been shown that, in the large $N$ limit, the
observables of that theory are given by a discrete unitary matrix model with
a certain potential $V(U)=T^{2}V_{2}\upsilon \left( U\right) $, where $T$ is
the temperature (inverse of the radius of $S^{1}$), $V_{2}$ is the volume of 
$S^{2}$ and $\upsilon \left( U\right) $ is the potential that has to be
computed, in an effective field theory approach, for every theory (that is,
depends on the choice of matter made) \cite{Jain:2013py}.

Thus, the matrix model description is precisely the one that holds for, more
realistic, \textit{finite} spin chains. Notice that the potential now is not 
$\ln \theta _{3}(\mathrm{e}^{i\varphi },q)$ in general. This simply means
that other choices for the coefficients $a_{k}$ should be made, but, as we
have seen, any system characterized by a well-defined unitary matrix model
with single trace terms $c_{k}\mathrm{Tr}U^{k}$ with $k\in 
\mathbb{Z}
$ in the potential is described in the same manner by the corresponding spin
chain model.

It is also interesting to explore how robust are the quantum topological
properties of the spin chain. Namely, is the interaction chosen in (\ref%
{CS-H}) unique ? In particular, for the Chern-Simons unitary ensemble, it is
known that $g(\varphi )=\theta _{3}^{-1}(-\mathrm{e}^{i\varphi },q)$ is also
a valid weight function for the matrix model to describe Chern-Simons theory 
\cite{Szabo:2010sd}. In our context, this implies that we can modify the
interactions in (\ref{CS-H}) with an alternating
ferromagnetic-antiferromagnetic coupling. If one considers, as above, the
Fourier coefficients of $\ln g(\varphi )=-\ln \theta _{3}(-\mathrm{e}%
^{i\varphi },q),$ then the corresponding Fourier coefficients (\ref{decay})
are modified accordingly to%
\begin{equation}
b_{k}=\frac{1}{2k\sinh \left( \frac{k\gamma }{2}\right) },  \label{decay2}
\end{equation}%
and the resulting spin chain%
\begin{equation}
\hat{H}_{CS}=-\sum_{i,k\in 
\mathbb{Z}
_{+}}\frac{1}{2k\sinh \left( \frac{k\gamma }{2}\right) }\left( \sigma
_{i}^{-}\otimes \sigma _{i+k}^{+}+\sigma _{i}^{-}\otimes \sigma
_{i-k}^{+}\right) +\frac{h}{2}\sum_{i}(\sigma _{i}^{z}-\mathbb{I}),
\label{HCS}
\end{equation}%
possesses the same topological properties. Notice that, in the expression
given above (Main result 2) we are led to an alternating
ferromagnetic-antiferromagnetic interaction but with the same interaction
coefficients otherwise. However, the first and leading interaction is in
both cases ferromagnetic. This equivalence of both spin chain models is
exact for the partition function $S_{00}$ \cite{Szabo:2010sd}, which is
given by a matrix model without any insertion of a Schur polynomial.
Regarding the quantum dimensions or the full topological S-matrix, the
alternative choice (\ref{decay2}) gives the same observables but with
transposed partitions $\lambda ^{\prime },\mu ^{\prime }$ (the columns of
the Young tableaux becomes rows and conversely). Thus, alternatively, if one
wants equality of the correlators the respective pattern of flipped spins is
different (but immediately related) depending on the election of (\ref{decay}%
) or (\ref{decay2}).

Notice that there is also freedom in the choice of the external magnetic
field since Chern-Simons theory is actually a theory of \textit{framed}
knots and links. Framing $\Pi $ on a three-manifold $M$ (here $M=S^{3}$), is
a choice of trivialization of the bundle $TM\oplus TM$ \cite{Atiyah}. It
contributes multiplicatively to the observables and for gauge group $G$ its
explicit expression is \cite{ST}%
\begin{equation}
\delta (M,\Pi )=\exp \Big(\frac{2\pi {\,\mathrm{i}\,}s}{24}\,c\Big)=\exp %
\Big(\frac{\pi {\,\mathrm{i}\,}s(N^{2}-1)k}{\,(k+N)}\Big)\ ,
\label{framJeffrey}
\end{equation}%
where $c$ is the central charge of the associated Wess-Zumino-Witten model
based on the affine extension of $G$. Notice that it is parametrized by an
integer $s\in \mathbb{Z}$. In the second expression we have particularized
for the case $G=U(N)$. At the level of anyon physics, the framing
contribution in the Chern-Simons theory describes how anyons rotate while
they wind around each other \cite{F}.

For example with the choice $h=\sum_{n=1}^{\infty }\ln (1-q^{n})$ to account
for the zeroth order Fourier coefficients of the potential, the correlator
is exactly the theta function matrix model, whose partition function is
exactly \cite{RT} 
\begin{equation*}
\,\langle ...,\uparrow ,\underset{N}{\underbrace{\downarrow ,...,\downarrow }%
}\rvert {}\mathrm{e}^{-\hat{H}_{CS}}\,\lvert \underset{N}{\underbrace{%
\downarrow ,...,\downarrow }},\uparrow ,...\rangle =\mathrm{e}^{-i\pi
N(N+1)}\prod\limits_{i=1}^{N-1}(1-q^{i})^{N-i}.
\end{equation*}%
To account for the framing contribution (\ref{framJeffrey}) one can simply
add the corresponding term in the magnetic field $h$ taking into account the
general result (\ref{matb}).

\section{Open problems}

There are several interesting open questions. First, to what extent finite
spin chains can be engineered to reproduce the observables of a Chern-Simons
theory with matter. This theory is no longer topological and the degeneracy
at zero energy of Chern-Simons theory is broken. It would be interesting to
understand if the finite size of the spin chain model achieves this in a
natural way. Second, we have seen that the XX models extended with
additional interactions have remarkable properties from the point of view of
the theory of exactly solvable systems since its correlation functions have
an immediate connection with the WZW model and admit very well-known exact
expressions for every $N$, something that does not hold in the case of the
ordinary XX model. Thus, it would be interesting now to study the properties
of the Hamiltonian, including spectral properties and the establishment of
its universality class, which is expected to be the same as that of the XX
model, due to the very fast decay of the additional interactions. Comparison
with other exactly solvable spin chains with infinitely many interactions
such as the Inozemtsev model \cite{Ino} or the Richardson-Gaudin models
would be also of interest.

To extend the discussion to the case of $q$ root of unity is another
interesting open problem. The type of series expansions involved in the
discussion of Chern-Simons theory are $q$-series, which happen to be much
better known for $q$ real and $q<1$ (or $q>1$). Indeed, these series are
analytic inside $\left\vert q\right\vert <1$ and they have $\left\vert
q\right\vert =1$ as a natural boundary. Notice for example also that for $q$
root of unity the power series coefficients above are not even defined, due
to the appearance of terms such as the $q$-Pochhammer symbol $\left(
q;q\right) _{n}:=\prod\nolimits_{j=0}^{n-1}(1-q^{j})$. In spite of this, it
can be shown that most properties holds in that case as well \cite{DSL}, and
would be interesting to take these results into account to reproduce the
construction here for $q$ root of unity. Furthermore, these more delicate
analysis of the $q$-series may lead to further understanding of the model.
For example, a well-known identity%
\begin{equation}
\sum_{n=0}\frac{z^{n}}{\left( q;q\right) _{n}}=\exp \left(
\sum_{n=1}^{\infty }\frac{z^{n}}{n(1-q^{n})}\right) ,  \label{HL}
\end{equation}%
was shown by Hardy and Littlewood to hold, even for $q$ root of unity \cite%
{DSL}. One could then use (\ref{HL}) and similar properties \cite{DSL} to
extend our result to $q$ roots of unity.

However, this needs to be thoroughly investigated, both for $q$ real and $q$
root of unity. Another clear open problem is to extend this description to
the case of knot and link polynomials invariants. Very recently it has been
shown that a number of colored HOMFLY polynomial invariants can be written
as a terminating basic hypergeometric function \cite{Tiknot}. A possible way
to relate this to the results presented here is the fact that these
functions admit an integral representation, which in turn can be also, in
principle, identified with a unitary matrix model. This is left as an open
problem, to be discussed elsewhere.

\subsection*{Acknowledgements}


We thank Bel\'{e}n Paredes and Germ\'{a}n Sierra for discussions at initial
stages of this work and Milosz Panfil for comments. The work of MT was
concluded while starting as Investigador FCT at Universidade de Lisboa. Both
authors acknowledge support from MINECO (grant MTM2011-26912), Comunidad de
Madrid (grant QUITEMAD+-CM, ref. S2013/ICE-2801) and the European CHIST-ERA
project CQC (funded partially by MINECO grant PRI-PIMCHI-2011-1071). This
work was also made possible through the support of grant \#48322 from the
John Templeton Foundation. The results obtained in this publication are
those of the authors and do not necessarily reflect the views of the John
Templeton Foundation.

\appendix

\section{Definitions}

We introduce some Lie algebra conventions and definitions, focussing on $%
U(N) $. We use the standard notations as in \cite{CFT} $\omega _{i}$ for $%
i=1,\ldots ,N$, denotes the set of fundamental weights, and ${\,\mathrm{e}}%
_{i}$ are unit vectors in ${\mathbb{R}}^{N}$. The set of fundamental weights
is then defined by%
\begin{equation}
\omega _{i}=\sum_{j=1}^{i}{\,\mathrm{e}}\,_{j}-{\frac{i}{N}}\sum_{j=1}^{N}{\,%
\mathrm{e}}\,_{j}.
\end{equation}%
There is also the notion of simple roots, which are given by $\alpha _{i}={\,%
\mathrm{e}}\,_{i}-{\,\mathrm{e}}\,_{i+1}$ for $i=1,\ldots ,N-1$. One can
expand the highest weight of an irreducible representation%
\begin{equation}
\lambda =\sum_{i=1}^{N}\lambda _{i}\omega _{i}=\sum_{i=1}^{N}(\ell
_{i}-\kappa ){\mathrm{e}}_{i}  \label{loi}
\end{equation}%
where%
\begin{equation}
\kappa ={\frac{1}{N}}\sum_{j=1}^{N}j\ell _{j}
\end{equation}%
is the number of boxes in the Young tableaux divided by $N$, and 
\begin{equation}
\ell _{i}=\lambda _{i}+\lambda _{i+1}+...+\lambda _{i+N}.
\end{equation}%
With these definitions, it is immediate to check that (\ref{loi}) holds. The
Weyl chamber is a connected set in ${\mathbb{R}}^{N}$ left out after we
delete all hyperplanes orthogonal to the roots. Its definition is%
\begin{equation*}
C_{\omega }=\left\{ \lambda \left( \omega \lambda ,\alpha _{i}\right) \geq 0,%
\text{ }i=1,...N-1\right\} ,\text{ }\omega \in W,
\end{equation*}%
where $W$ denotes the Weyl group. The fundamental chamber mentioned in the
Introduction corresponds to the identity element, $C_{0}$. It is manifest
from the definition that $\lambda _{i}\geq 0$ if $\lambda $ is in the
fundamental chamber and, therefore, the $\ell ^{\prime }s$ satisfy $\ell
_{i}\geq \ell _{i+1}$ and one can define the coordinates:%
\begin{equation*}
h_{i}=\ell _{i}+\frac{N+1}{2}-i.
\end{equation*}%
The fundamental chamber then is the domain $h_{1}>h_{2}>...>h_{N}$. These
are the variables that are usually used to label representations, and which
we have labeled by $\lambda $. They can be written as%
\begin{equation*}
h=\ell +\rho
\end{equation*}%
where $\rho $ is the Weyl vector $\rho :=\sum_{i=1}^{N}\omega _{i}$

This explains the notations in the Introduction and how the states of the
Hilbert space, denoted by $\left\vert \lambda \right\rangle ,$ and which are
integrable representations, can be indexed, at the practical level, by Young
tableaux.



\end{document}